\documentclass[12pt,letterpaper]{article}         
\usepackage{jheppub}
\usepackage{epsfig}

\def\be{\begin{eqnarray}}
\def\ee{\end{eqnarray}}

\newcommand\para{\paragraph{}}

\newcommand{\eqn}[1]{(\ref{#1})}

\def\Dslash{\,\,{\raise.15ex\hbox{/}\mkern-12mu D}}
\def\Dbarslash{\,\,{\raise.15ex\hbox{/}\mkern-12mu {\bar D}}}
\def\delslash{\,\,{\raise.15ex\hbox{/}\mkern-9mu \partial}}
\def\delbarslash{\,\,{\raise.15ex\hbox{/}\mkern-9mu {\bar\partial}}}
\def\pslash{\,\,{\raise.15ex\hbox{/}\mkern-9mu p}}
\def\calDslash{\,\,{\raise.15ex\hbox{/}\mkern-12mu {\cal D}}}

\newcommand{\RN}{Reissner-Nordstr\"om}

\def\lae{\mathrel{\mathop{\smash{\lower .5 ex \hbox{$\stackrel<\sim$}}}}}
\def\lae{\mathrel{\mathop{\smash{\lower .5 ex \hbox{$\stackrel>\sim$}}}}}

\preprint{DAMTP-2015-37}

\title{\large Magnetotransport from the fluid/gravity correspondence}

\author{Mike Blake} 

\affiliation{ Department of Applied Mathematics and Theoretical Physics,
 University of Cambridge, Cambridge, CB3 OWA, UK}

\emailAdd{m.a.blake@damtp.cam.ac.uk}

\abstract{We continue our construction of a hydrodynamical description of a holographic model with broken translation invariance. Using the fluid/gravity correspondence we derive the constitutive relations of the boundary theory in the presence of a magnetic field. This allows us to obtain novel results for the low-frequency magnetothermoelectric response coefficients.  We discuss the DC limit of our hydrodynamics in detail, and show that our approach is equivalent to the `horizon-fluid'  of Donos and Gauntlett. }

\begin{document}
\maketitle
\pagestyle{plain} \setcounter{page}{1}
\newcounter{bean}
\baselineskip16pt

\section{Introduction}

\para{}In the last few years there has been a resurgence of interest in studying heat and charge transport in holographic theories. By finding ways to incorporate momentum relaxation into these models \cite{jorge1, jorge2, sandiego, vegh, davison, andradewithers, Qlattices} rapid progress has been made in understanding their DC \cite{univdc, lattices, aristosdc, newblaise1, newblaise2, thermo,amoretti1}, and low-frequency \cite{lucasac, richblaise1, nernst, lucasmagneto, hydrous} response coefficients. 
\para{}Perhaps the most interesting aspects of these results are that holography provides, beyond leading order in the strength of momentum relaxation, a class of non-Drude models of transport (see for instance \cite{metalinsulator, aristosdc, newblaise1, richblaise1,richblaise2, kimincoh}). Such models are likely to be necessary to describe the experimental measurements on strange metals \cite{incoherent, ong, hall}.  
 \para{} It is therefore an important open problem to develop a detailed framework in which to study this generalised transport. Motivated by this goal, we recently utilised the fluid/gravity correspondence \cite{fluidgrav, hubeny, chargedhydro,erdmenger, minwalla} to construct a hydrodynamical description of the boundary physics dual to a simple holographic theory with broken translational invariance \cite{hydrous}. More precisely, we studied a 4+1 dimensional Einstein-Maxwell-Dilaton theory in which the translational symmetry was broken by linear sources $\phi^{(0)}_{\cal A} = k x_{\cal A}$ for scalar fields dual to marginal operators ${O}_{\cal A}$.
\para{}The resulting hydrodynamical description then consists of the constitutive relations, which express the electrical and heat currents of the boundary theory in terms of a local fluid velocity $u^{\mu}(x)$, together with the Ward-identity that describes the relaxation of momentum by the scalars. It was found that, beyond leading order in the derivative expansion, both the constitutive relations and Ward identity differed from those studied in the seminal work of Hartnoll et al \cite{nernst}. Nevertheless, the resulting hydrodynamics was remarkably simple and could be used to calculate the thermoelectric response coefficients of the boundary theory. 
\para{}In this paper, we continue to develop this approach by studying the magnetohydrodynamics dual to the 3+1 dimensional version of the model studied in \cite{hydrous}.  The effect of the magnetic field is to introduce new terms into the constitutive relations and to modify that Ward identity to include the Lorentz force
\be
\partial_{\mu} T^{\mu \nu} = \partial^{\nu} \phi^{(0)}_{\cal A} \langle O_{\cal A} \rangle + {\cal F}^{\nu \mu} J_{\mu} 
\label{wardidentity}
\ee
Here ${\cal F}^{\mu \nu}$ is the field strength of an external gauge field that we use to turn on a magnetic field, $B$. Using the fluid/gravity correspondence we evaluate the constitutive relations to ${\cal O}(\varepsilon^2)$ in our derivative expansion and the Ward identity to ${\cal O}(\varepsilon^4)$. Once again we find that whilst the results agree with \cite{nernst} at leading order, there are subleading corrections that need to be taken into account.  
\para{}Given the initial motivation of \cite{nernst} with relation to the Nernst effect, it is particularly important to understand how these corrections effect the magnetotransport of the boundary theory. By linearising our constitutive relations in the fluid velocity, $v_i$, we are able to calculate new results for the entire set of low-frequency thermoelectric response coefficients.
\para{}Of special interest is the $\omega \rightarrow 0$ limit of these results. It has long been known that this limit is very special within holographic models - in particular it is possible to obtain exact expressions for the DC response coefficients in terms of horizon data \cite{hall, magnetous, magnetoitaly, magnetokorea}. We therefore end this paper by reformulating our constitutive relations in a new hydrodynamical frame in order to make the structure behind these DC formulae self-evident. In the DC limit, this approach is found to be equivalent to the exact `horizon-fluid' recently proposed by Donos and Gauntlett \cite{hydroaristos,hydroaristos2}. 
\para{}The remainder of this paper is organised as follows. In Section~\ref{sec:fluidgravity} we use the fluid/gravity correspondence to construct the constitutive relations of magnetohydrodynamics dual to our holographic model. Since much of the discussion is equivalent to that of \cite{hydrous}, we will be schematic in our presentation of the details. In Section~\ref{sec:linearhydro} we study the linear response of the boundary theory and extract the thermoelectric response coefficients. Finally in Section~\ref{sec:dclimit} we focus on the DC limit and examine the connection with \cite{hydroaristos,hydroaristos2}.    
\section{The fluid/gravity correspondence} 
\label{sec:fluidgravity}
 \paragraph{}In this section we will explain how to derive the constitutive relations of the magnetohydrodynamics dual to a simple holographic model with broken translational invariance. Of particular phenomenological interest \cite{nernst, miraculous1, miraculous2} are 2+1 dimensional boundary CFTs, and so we will use the four-dimensional Einstein-Maxwell action in the bulk
\be
 S = \frac{1}{16 \pi G_N}\int \mathrm{d}^4 x \sqrt{-g}\bigg[ R + 6 - \frac{1}{4} F^{MN}F_{MN} - \frac{1}{2} g^{MN} \partial_{M} \phi_{\cal A} \partial_{N} \phi_{\cal A} \bigg ] 
 \label{action}
\ee
where the scalar fields $\phi_{\cal A}$ will be used to introduce momentum relaxation into the boundary theory. Here the calligraphic index on the scalars runs over the spatial directions on the boundary, i.e. ${\cal A} = 1, 2$. Note that in \cite{hydrous} we recently studied the 5 dimensional analogue of this theory (without the magnetic field). Since we can always set $B=0$, as a byproduct of our calculations we will generalise the results of \cite{hydrous} to 2+1 dimensional boundary theories. 
 \para{}The starting point in constructing the boundary hydrodynamics is the \RN\ black brane in ingoing Eddington-Finkelstein coordinates
\begin{eqnarray}
g_{M N} dx^{M} dx^{N} &=& - 2 u_{\mu} dx^{\mu} dr - r^2 f(r) u_{\mu} u_{\nu} dx^{\mu} dx^{\nu} + r^2 P_{\mu \nu} dx^{\mu} dx^{\nu} \nonumber \\
A_M dx^{M} &=&  \frac{2 q}{r} u_{\mu} dx^{\mu}  \nonumber \\
\phi_{\cal A} &=& \phi_0
\label{blackhole}
\end{eqnarray}
Where $x^{\mu}$ are the boundary coordinates, $u_{\mu} dx^{\mu} = -dv$ and $P_{\mu \nu} = \eta_{\mu \nu} + u_{\mu} u_{\nu}$ is the projector perpendicular to $u_{\mu}$. In four bulk dimensions the emblackening factor is given by 
\be
f(r) = 1 - b/r^3 + q^2/r^4
\ee 
and the  horizon radius, $r_0$, is defined as the solution to 
\be
b = r_0^3\bigg( 1 + \frac{q^2}{r_0^4} \bigg)
\ee
We stress that, although our primary interest in this paper is magnetohydrodynamics, the Reisnner-Nordstrom black brane in \eqn{blackhole} only contains an electric charge. The magnetic field, $B$, will be included in our description as part of the derivative expansion. As a result, the hydrodynamics we derive will be perturbative in the strength of the magnetic field\footnote{For other early approaches to incorporating a magnetic field within fluid-gravity see \cite{oscar, kraus}.}. 
\subsection*{The derivative expansion}
\paragraph{}To derive the constitutive relations of the boundary theory we follow the standard proceedure of the fluid gravity correspondence (see for instance \cite{fluidgrav, hubeny,minwalla,chargedhydro,erdmenger}). First we take the parameters $u^{\mu}, q, b, \phi_{\cal A}$ appearing in \eqn{blackhole} to be functions of boundary co-ordinates $x^{\alpha}$. Following \cite{son, chiralmag, external} we allow for an external gauge field $A^{\mathrm{ext}}_{\mu}(x^{\alpha})$ that we will use to apply a magnetic field to the boundary theory. We therefore take the ansatz 
\be
g^{(0)}_{M N} dx^{M} dx^{N}   &=& - 2 u_{\mu}(x^{\alpha}) dx^{\mu} dr - r^2 f(r, q(x^{\alpha}),b(x^{\alpha})) u_{\mu}(x^{\alpha}) u_{\nu}(x^{\alpha}) dx^{\mu} dx^{\nu} + r^2 P_{\mu \nu}(x^{\alpha}) dx^{\mu} dx^{\nu} \nonumber \\
A_M^{(0)}dx^{M} &=&  \frac{2 q(x^{\alpha})}{r} u_{\mu}(x^{\alpha}) dx^{\mu} + A^{\mathrm{ext}}_{\mu}(x^{\alpha}) dx^{\mu} \nonumber \\
\phi_{\cal A}^{(0)} &=& \phi^{(0)}_{\cal A}(x^{\alpha})
\label{ansatz}
\ee
The interpretation of this prescription is to think of these parameters $q(x^{\alpha}), b(x^{\alpha}), u^{\mu}(x^{\alpha})$ as corresponding to a local charge density, energy density and fluid velocity. In contrast, the scalars $\phi^{(0)}_{\cal A}(x^{\alpha})$ and gauge field $A^{\mathrm{ext}}_{\mu}(x^{\alpha})$ are external sources in the boundary theory. 
\para{}At zeroth order, i.e. when these fields take constant values, this ansatz satisfies the Einstein-Maxwell-Dilaton equations. More generally, we need to supplement \eqn{ansatz} with corrections that can be calculated by solving the Einstein-Maxwell-Dilaton equations order by order in a derivative expansion. After determining these corrections, the expectation value of any operator in the boundary theory can be extracted in terms of the hydrodynamic fields, together with an expansion in the derivatives of these variables and the external sources\footnote{Of course, an expansion of derivatives only makes sense if the derivative are small in comparison to the scales set by the background theory. That is most simply achieved by taking the derivatives to be small with respect to the chemical potential $\mu$ and temperature $T$. }  $A^{\mathrm{ext}}_{\mu}, \phi_{\cal A}^{(0)}$.
\paragraph{}In this paper we are interested in studying solutions in which the external sources $\phi^{(0)}_{\cal A}$ and $A_{\mu}^{\mathrm{ext}}$ take specific forms. Firstly, we will break translational invariance using the sources
\be
\phi_{1}^{(0)} = k x \;\;\;\; \phi_{2}^{(0)} = k y
\label{scalarsources}
\ee
These sources, first studied in the context of transport in \cite{andradewithers}, are the simplest way we can incorporate momentum relaxation within our holographic model. Similarly, we will use the external gauge field to introduce a magnetic field to the boundary field
\be
A^{\mathrm{ext}}_y = B x
\label{gaugesources}
\ee
Note that the choice of sources \eqn{scalarsources} and \eqn{gaugesources} defines a laboratory frame $(t , x, y)$. 
\paragraph{}Our goal in this paper is to study how the breaking of translation invariance and the introduction of the magnetic field affect the thermoelectric transport properties of the boundary theory. Since the bulk stress tensor is quadratic in derivatives of the scalar fields, we will find that at leading order momentum relaxes at a rate $\tau^{-1} \sim k^2$. We therefore take the anisotropic scalings
\be
k \sim \varepsilon \;\;\;\;\; B \sim \varepsilon^2 \;\;\;\; \partial_{\mu} q \sim \varepsilon^2, \;\;\;\; \partial_{\mu} b \sim \varepsilon^2 \;\;\;\  \partial_{\mu} u^{\alpha} \sim \varepsilon^2 \;\;\;\; 
\ee
so that the frequency $\omega$ of the fluid flow, the relaxation rate $\tau^{-1}$, and the cyclotron frequency $\omega_c \sim B$ all have the same scaling.  Note that, once we go beyond leading order in this expansion, the equilibrium configuration is no longer the translationally invariant black brane \eqn{blackhole}. Rather, the equilibrium solution changes in the presence of the sources \eqn{scalarsources} and \eqn{gaugesources}. These corrections can be determined order by order within our expansion by setting $u^{\mu} = (1,0,0)$ and $ \partial_{\mu} q = \partial_{\mu} b = 0$, but retaining the terms proportional to derivatives of $A^{\mathrm{ext}}_{\mu}$ and $\phi^{(0)}_{\cal A}$ . 
\para{} We can now perform our derivative expansion by solving the bulk equation of motions as a perturbation series in $\varepsilon$. To study transport in the boundary theory, we need to evaluate $J^{\mu}$ and $T^{\mu \nu}$ to ${\cal O}(\varepsilon^2)$ and to calculate $\langle O_{\cal A} \rangle$ up to\footnote{Note that to simplify our construction we will neglect certain terms that will not appear in the linearized hydrodynamics we are ultimately interested in. For our choice of sources \eqn{scalarsources}, this means that we will consistently ignore any terms in the constitutive relations proportional to 
$u^{\mu} u^{\nu} \partial_{\mu} \phi^{(0)}_{\cal A} \partial_{\nu} \phi^{(0)}_{\cal B}$.} ${\cal O}(\varepsilon^3)$. The corresponding calculation for the five dimensional analogue of this model was recently explained in detail in \cite{hydrous}. It is straightforward to adapt this computation to four dimensions and also to include the external gauge field\footnote{The only place the external field strength ${\cal F}_{\mu \nu}$ enters the equations of motions is through the vector-channel constraint equation. That is the 3+1 dimensional analogue of equation (A.18) in \cite{hydrous}.} \cite{son,chiralmag,external}. We therefore simply present our final results for constitutive relations, and refer the interested reader to this literature. 
\subsection*{Thermodynamics}
\paragraph{}Before proceeding to discuss hydrodynamics, we should quickly review the thermodynamics of the boundary theory. As we remarked above, the equilibrium solutions, and hence thermodynamics, receive corrections in the presence of the scalar sources and magnetic field. Fortunately, since we are fixing the local charge and energy densities in our expansion, these take the same form as in the black hole background \eqn{blackhole} 
\be
\epsilon = \frac{2 b}{16 \pi G_N} \;\;\; \rho = \frac{2 q}{16 \pi G_N}
\ee
Conversely the remaining thermodynamics variables $s, T, \mu$ and $P$ are corrected at ${\cal O}(\varepsilon^2)$ by the presence of the scalar fields \eqn{scalarsources}. Due to the scaling $B \sim \varepsilon^2$, we do not see the corrections in the thermodynamics due to the magnetic field at this order. 
\paragraph{}We find that the first-order corrected entropy density is given by
\be
s = \frac{1}{4 G_N} \bigg(r_0^2 + \frac{k^2}{3 - Q^2}\bigg) + \dots 
\ee
where $Q = q/r_0^2$ is the dimensionless charge density, and $\dots$ indicates terms of higher order in our expansion. We can also calculate the chemical potential
\be
\mu &=& \frac{2 q}{r_0} - \frac{q k^2}{r_0^3(3 - Q^2)} + \dots 
\ee
the temperature
\be
4 \pi T &=& r_0(3 - Q^2) + \frac{2 Q^2 k^2}{r_0(3 - Q^2)} + \dots 
\ee
and finally the pressure 
\be
P &=& \mu \rho + s T - \epsilon = \frac{1}{16 \pi G_N} \bigg( b + k^2 r_0 \bigg) + \dots
\ee
\subsection*{Constitutive relations}
\paragraph{}Having determined the thermodynamics we can now consider the constitutive relations. To make these well-defined, we first need to pick a fluid-dynamical frame. This choice reflects the fact that, out of equilibrium, there is an ambiguity in defining the local fluid velocity. In the study of finite-density hydrodynamics, it is conventional to choose the Landau frame condition
\be
u_{\mu} T^{(1) \mu \nu} = 0
\label{landau}
\ee
where $T^{(1) \mu \nu}$ is the first-order correction to the stress tensor. Physically this condition defines $u^{\mu}$ to be the velocity of the energy current\footnote{Note, as will be crucial in our discussion of Section~\ref{sec:dclimit}, this is distinct from the velocity of the heat current.}. 
\paragraph{}With this condition the constitutive relation for the stress tensor can be written as
\be
T^{\mu \nu} &=& T^{(0){\mu \nu}} + T^{(1) \mu \nu} 
\ee
where 
\begin{eqnarray}
T^{(0) {\mu \nu}} &=& \frac{2 b}{16 \pi G_N} u^{\mu} u^{\nu} + \frac{b}{16 \pi G_N} P^{\mu \nu}  \nonumber \\
T^{(1){\mu \nu}} &=&  - \frac{2 r_0^2}{16 \pi G_N} \sigma^{\mu \nu} - \frac{r_0}{16 \pi G_N} \Phi^{\mu \nu}
\label{stressconstit}
\end{eqnarray}
and the tensorial sources appearing in the first order correction are 
\begin{eqnarray}
\sigma^{\mu \nu} &=& P^{\mu \alpha} P^{\nu \beta} \partial_{(\alpha} u_{\beta)} - \frac{1}{2} P^{\mu \nu} \partial_{\alpha}u^{\alpha} \nonumber \\
\Phi^{\mu \nu} &=& P^{\mu \alpha} P^{\nu \beta} \partial_{\alpha} \phi^{(0)}_{\cal A} \partial_{\beta} \phi^{(0)}_{\cal A} - \frac{1}{2} P^{\mu \nu} P^{\alpha \beta} \partial_{\alpha} \phi^{(0)}_{\cal A} \partial_{\beta} \phi^{(0)}_{\cal A} 
\end{eqnarray}
Likewise the electrical current can be extracted as
\begin{eqnarray}
J^{\mu} &=& \frac{2 q}{16 \pi G_N} u^{\mu}  - \frac{2}{16 \pi G_N} \frac{M + 2}{3 M r_0} P^{\mu \nu} (\partial_\nu q - \frac{2 q}{3 b} \partial_{\nu} b)  \nonumber \\  &+&  \frac{1}{16 \pi G_N} \bigg( \frac{3 - Q^2}{3 M} \bigg)^2P^{\mu \nu} {\cal F}_{\nu \lambda} u^{\lambda} - \frac{1}{16 \pi G_N} \frac{8 Q^3}{9 M^2} (u^{\lambda} \partial_{\lambda} {\phi}^{(0)}_{\cal A}) P^{ \mu \nu} \partial_{\nu} \phi^{(0)}_{\cal A}
\label{currentconstit}
\end{eqnarray}
where $M = b/r_0^3$ and ${\cal F}_{\mu \nu}$ is the field strength of the external gauge field $A^{\mathrm{ext}}_{\mu}$. 
\paragraph{}Whilst this expression appears somewhat ungainly, we can make the structure much more evident by trading derivatives of $q, b$ with the conjugate thermodynamics variables $\mu, T$. This yields
\begin{eqnarray}
J^{\mu} = \rho u^{\mu}  + \sigma_Q  P^{\mu \nu} \bigg(- \partial_{\nu} \mu + {\cal F}_{\nu \lambda} u^{\lambda} + \mu \frac{\partial_{\nu} T}{T}  \bigg)  - \frac{1}{16 \pi G_N} \frac{8 Q^3}{9 M^2} (u^{\lambda} \partial_{\lambda} {\phi}^{(0)}_{\cal A}) P^{ \mu \nu} \partial_{\nu} \phi^{(0)}_{\cal A} \nonumber \\
\label{chargemu}
\end{eqnarray} 
where it is now clear that the external field strength and chemical potential enter the constitutive relations in the natural combination $-\partial_{\nu} \mu + {\cal F}_{\nu \lambda} u^{\lambda}$. These terms are proportional to a single transport coefficient
\be
\sigma_Q = \frac{1}{16 \pi G_N} \bigg( \frac{s T}{\epsilon + P} \bigg)^2
\label{sigmaQ}
\ee
The first two terms in \eqn{chargemu} are precisely the constitutive relation of relativistic hydrodynamics \cite{kovtun, nernst} with the specific choice \eqn{sigmaQ} for $\sigma_Q$. However, as was first appreciated in \cite{hydrous}, the presence of the scalars $\phi_{\cal A}$ breaks Lorentz invariance and introduces new terms into the constitutive relation that are absent in \cite{nernst}. 
\subsection*{Scalar expectation value}
Finally we need the constitutive relation for the scalar up to ${\cal O} (\varepsilon^3)$. The calculation although somewhat involved, is identical to the one performed in 5 dimensions in \cite{hydrous}. For our purposes, all we need to know is that we can expand $\langle O_{\cal A} \rangle$ as
\begin{eqnarray}
\langle O_{\cal A} \rangle =&-& \frac{r_0^2}{16 \pi G_N} u^{\lambda} \partial_{\lambda} \phi^{(0)}_{\cal A} - \frac{\lambda r_0}{16 \pi G_N} {\cal S}^+_{\cal A} + \frac{r_0}{16 \pi G_N} {\cal S}^-_{\cal A} \nonumber \\ &-& \frac{1}{16 \pi G_N} \frac{2}{4 (3 -Q^2)} u^{\lambda} \partial_{\lambda} \phi^{(0)}_{\cal A} (P^{\mu \nu} \partial_{\mu} \phi^{(0)}_{\cal B} \partial_{\nu} \phi^{(0)}_{\cal B}) \nonumber \\ &+& \frac{1}{16 \pi G_N} \frac{2}{3 M} u^{\lambda} \partial_{\lambda} \phi^{(0)}_{\cal B} (P^{\mu \nu} \partial_{\mu} \phi^{(0)}_{\cal A} \partial_{\nu} \phi^{(0)}_{\cal B}) \nonumber \\ &-&\frac{1}{16 \pi G_N} \frac{2 q P^{\mu \nu} \partial_{\mu} \phi^{(0)}_{\cal A}( \partial_{\nu} q + 2 q u^{\lambda} \partial_{\lambda} u_{\nu})}{3 M r_0^3} 
\label{secondscalarvev}
\end{eqnarray}
where the source terms ${\cal S}^{\pm}_{\cal A}$ are defined to be
\begin{eqnarray}
{\cal S}^{\pm}_{\cal A} &=& (u^{\lambda} \partial_{\lambda} u^{\mu} \pm \frac{1}{2}\partial_{\lambda}u^{\lambda} u^{\mu}) \partial_{\mu} \phi^{(0)}_{\cal A} 
\end{eqnarray}
Note that, written this way, there is no explicit dependence on the external field strength ${\cal F}^{\mu \nu}$ appearing in this constitutive relation. 
\para{}The only unspecified quantity is then the transport coefficient $\lambda$, that multiplies ${\cal S}^{+}_{\cal A}$ in \eqn{secondscalarvev}. In general this is a complicated function of $\mu/T$ that we have not been able to determine analytically. Nevertheless, it can be calculated perturbatively in $Q$ for which we find the leading terms
\be
\lambda = -\frac{\sqrt{3} \pi - 9 \mathrm{log}3}{18} - 
 \frac{5 \sqrt{3} \pi -72 + 27 \mathrm{log}3}{54} Q^2 + \dots
 \label{lambda}
\ee
Whilst this is a rather complicated expression, the precise value of $\lambda$ will not be important in describing the magnetotransport of the boundary theory - which we study in detail in the next section.
\section{Thermoelectric response coefficients}
\label{sec:linearhydro}
In the last section we quickly reviewed how the fluid-gravity correspondence can be used to derived the constitutive relations of the boundary theory. In the remainder of this paper, we wish to use this hydrodynamics to study how the magnetic field affects charge and heat transport. To do this, we need to linearise our constitutive relations around equilibrium. We therefore consider the fluid flow\footnote{Here $i=1,2$ are spatial indices, i.e. $x$, $y$ in the laboratory frame.}
\be
u_{t}  = -1 \;\;\; u_i = \bigg(1+ \frac{k^2 r_0}{3b}\bigg) v_i(x,y,t)
\label{fluidvel}
\ee
together with the perturbations
\begin{eqnarray}
\mu(x, y, t) \rightarrow  \mu + \delta \mu(x, y, t) \nonumber \\
T(x, y, t)  \rightarrow T + \delta T(x, y, t) 
\end{eqnarray}
\para{}The unusual normalisation of the fluid velocity $v_i$ in \eqn{fluidvel} has been chosen so that, after linearising in the perturbations $(\delta \mu, \delta T, v_i)$, the constitutive relation for the momentum density $T^{0 i}$ takes the same form as in relativistic hydrodynamics
\be
T^{0 i} &=& (\epsilon + P)v_i + ....
\ee
where the \dots indicate higher order terms, i.e. those that are ${\cal O}(\varepsilon^4)$. The corresponding expression for the electrical current is
\be
J_i = \rho v_i + \sigma_{Q}\bigg(-\partial_{i} \mu + {\cal F}_{i j}v_{j} + \mu \frac{\partial_i T}{T} \bigg) + \frac{\mu \sigma_Q k^2}{4 \pi T} v_i+ \dots
\label{linearcurrent}
\ee
where ${\cal F}_{i j} = B \epsilon_{i j}$ is the field strength of the magnetic field \eqn{gaugesources}. Similarly, the heat current $Q^i = T^{0i} - \mu J^i$ has components
\be
Q_i = s T v_i - \mu \sigma_{Q}\bigg(-\partial_{i} \mu + {\cal F}_{i j} v_j + \mu \frac{\partial_i T}{T} \bigg) - \frac{\mu^2 \sigma_Q k^2}{4 \pi T} v_i \dots
\label{linearheat}
\ee
Note that the thermodynamic factors in these formulae include the ${\cal O}(\varepsilon^2)$ corrections we determined earlier. Written this way, we find it remarkable how simple these equations are. In the absence of the magnetic field, they take the same form as those derived from a five dimensional bulk action in \cite{hydrous}, although the precise value of $\sigma_Q$ is dimension dependent. 
\para{} Compared to the usual constitutive relations of relativistic hydrodynamics, there is just the one extra term ($\sim k^2$) due to the scalars. It is intriguing that, within these specific holographic models, this novel term is naturally proportional to the same transport coefficient $\sigma_Q$ that multiplies the derivatives of $\mu$ and $T$. It would certainly be interesting to understand if this continues to hold more generally. If so, it might suggest that there is some fundamental reason, such as the positivity of entropy production \cite{landaulifshitz,nernst}, for why there only appears to be a single transport coefficient at this order. 

\subsection*{Ward identity}
\paragraph{}In order to describe momentum relaxation we need to supplement these constitutive relations with the linearised Ward identity. To obtain this, we insert the constitutive relation for the scalar expectation values into \eqn{wardidentity} and then linearise the resulting expression in our perturbations. Since our aim in this paper is to calculate the zero-wavevector response coefficients, we will ignore terms proportional to spatial derivatives of the fluid velocity. At leading order the Ward identity then reads
\begin{eqnarray}
\partial_{t} T^{0 i}  + \partial_{i} P &=& -\frac{k^2 s}{4 \pi} v_i + {\cal F}_{i j}  J_j + \dots \nonumber \\
&=& -\frac{k^2 s}{4 \pi (\epsilon + P)} T^{0 i} + \frac{\rho B}{\epsilon + P} \epsilon_{i j} T^{0 j} + \dots
\label{drude}
\end{eqnarray}
and simply describes a Drude excitation with a momentum relaxation rate, $\tau^{-1}$, and cyclotron frequency, $\omega_c$, given by 
\be
\tau^{-1} = \frac{k^2 s}{4 \pi (\epsilon + P)} \;\;\;\; \omega_c = \frac{\rho B}{\epsilon + P} 
\label{drudepoles}
\ee
Evaluating the Ward identity at ${\cal O}(\varepsilon^4)$ is much more involved since it requires using the subleading terms in \eqn{secondscalarvev}. Amazingly, we find that it can be written in the compact form
\be
\partial_{t} T^{0 i} + \partial_{i} P &=&  -\frac{k^2}{4 \pi} \bigg[ s v_i - \frac{\mu \sigma_{Q}}{T}\bigg(-\partial_{i} \mu + {\cal F}_{i j} v_j + \mu \frac{ \partial_{i} T}{T} \bigg) -  \frac{\mu^2 \sigma_Q k^2}{4 \pi T^2} v_i \bigg] + {\cal F}_{i j} J_{j} -  \frac{\lambda k^2 r_0}{16 \pi G_N} \partial_{t} v_i \nonumber  \\
&=& -\frac{k^2}{4 \pi T} Q_i + {\cal F}_{i j} J_j -\frac{\lambda k^2 r_0}{16 \pi G_N} \partial_{t} v_i + \dots
\label{acward}
\ee
where, just as was noticed in the absence of the magnetic field in \cite{hydrous}, the contribution of the scalar fields gives rise to a term proportional to the heat current. In \cite{nernst} a similar form the Ward identity was studied\footnote{The time-dependent term proportional to $\lambda$ is also not present in their model.}, but instead with the momentum density $T^{0 i}$ appearing on the right hand side of \eqn{acward}. Since these two quantities differ in general, the magnetotransport of these holographic theories is not described by the results of \cite{nernst} once we go beyond the Drude limit. 
\paragraph{}In order to extract this physics, we use the Ward identity \eqn{acward} to solve for the fluid velocity $v_i$ in terms of $\partial_{i} \mu$ and $\partial_{i}T$. In the presence of a magnetic field, it is convenient to work with the complexified fields $v_\pm = v_x \pm i v_y$.  After a Laplace transform in time we then find an expression for the fluid velocity up to ${\cal O}(\varepsilon^2)$ as
\be
v_+ &=& -\frac{\bigg(\frac{4 \pi \rho}{k^2 s} + \frac{\mu \sigma_Q }{s T}  + \frac{\mu^2 \rho \sigma_Q }{s^2 T^2}\bigg) \tau^{-1} - \frac{i \sigma_Q B}{\epsilon + P}}{\tau^{-1} + \gamma - i (\omega - \omega_c)} \partial_+ \mu 
 - \frac{\frac{4 \pi \tau^{-1}}{k^2} +  \frac{i \mu \sigma_Q B}{T (\epsilon + P)} }{\tau^{-1} + \gamma - i (\omega - \omega_c)} \partial_{+} T
\label{vplus}
\ee
where we have eliminated the pressure using the identity $\delta P = \rho \delta \mu + s \delta T$. An analogous expression for $v_-$ follows from making the replacement $B \rightarrow -B$ in \eqn{vplus}. 
\para{}The location of the poles in this fluid flow are described by the momentum relaxation rate\footnote{With $\lambda$ given by \eqn{lambda} this expression agrees with the momentum relaxation rate for this model derived in \cite{richblaise1}.}
\be
\tau^{-1} = \frac{k^2 s}{4 \pi (\epsilon + P)} \bigg[ 1 - \frac{\mu^2 \sigma_Q k^2}{4 \pi s T^2} - \frac{\lambda k^2 r_0 }{3 b_0} + \dots \bigg]
\label{correctedgamma}
\ee
and the cyclotron frequency
\be
\omega_c = \frac{\rho B}{\epsilon +P} \bigg[1 + \frac{2 \mu \sigma_Q k^2}{4 \pi \rho T} - \frac{\lambda k^2 r_0}{3 b_0} + \dots \bigg]
\label{correctedcyclo}
\ee
which can both be seen to contain subleading corrections to \eqn{drudepoles}. Additionally there is a novel effect at this order in hydrodynamics where we see a contribution 
\be
\gamma = \frac{\sigma_Q B^2}{\epsilon + P}
\ee
to momentum relaxation arising from the magnetic field. This is a relativistic phenomenon which, at least at weak coupling, is usually thought of as arising from collisions between particles and holes undergoing cyclotron orbits in opposite directions \cite{nernst}. 
\subsection*{Magnetotransport}
\para{}Armed with our expression for the fluid velocity we can now extract the thermoelectric response coefficients of the boundary. To do this we insert \eqn{vplus} into the constitutive relations for the currents. The Kadanoff-Martin prescription \cite{kadanoff, nernst} then tells us that the electrical conductivity, $\sigma_{+}$, the thermoelectric conductivity, $\alpha_{+}$, and heat conductivity, $\bar{\kappa}_+$ can be read off as\footnote{Note that at non-zero wavectors this prescription is more complicated due to the decay of initial perturbations.}
\[
 \Bigg( \;\; \begin{matrix} J_+ \\ Q_+ \end{matrix} \;\; \Bigg) =\Bigg( \;\; \begin{matrix} \sigma_+ \;&\; \alpha_+   \\ 
 T  \alpha_+ \;&\; \bar{\kappa}_+  
 \end{matrix} \;\; \Bigg) \Bigg( \;\; \begin{matrix} -\partial_{+} \mu \\ -\partial_{+} T \end{matrix} \;\; \Bigg) 
 \label{transportcoefficients}
\]
Recalling the scalings $\omega \sim B \sim k^2 \sim \varepsilon^2$ this approach allows us to calculate the response coefficients as a perturbation series in $\varepsilon$. We find that the electrical conductivity can be written as
\be
\sigma_{+}(\omega) = \frac{\frac{4 \pi \rho^2 \tau^{-1}}{k^2 s} + \sigma_0 \tau^{-1} - i \sigma_Q (\omega + \omega_c)}{\tau^{-1} + \gamma - i (\omega - \omega_c)} + \dots
\ee
where the $\dots$ are terms of ${\cal O}(\varepsilon^2)$ and the expressions for $\tau^{-1}$ and $\omega_c$ include the corrections in \eqn{correctedgamma} and \eqn{correctedcyclo}. Note that there is a subtle, but important, distinction between the parameters $\sigma_0$ and $\sigma_Q$
\be
\sigma_0 = \frac{1}{16 \pi G_N} \;\;\;\;\;\;\;\;\;\; \sigma_Q = \frac{1}{16 \pi G_N} \bigg( \frac{s T}{\epsilon + P} \bigg)^2 
\ee
which are appearing in the electrical conductivity. It is then straightforward to extract the usual components of the conductivity tensor as
\be
\sigma_{xx} &=& \frac{\sigma_{+} + \sigma_{-}}{2}  \;\;\;\;\;\; \sigma_{xy} = \frac{\sigma_{-} - \sigma_{+}}{2 i}
\ee
The resulting expressions for the low-frequency electrical conductivities can then be written as
\begin{eqnarray}
\sigma_{xx}(\omega) &=& \frac{\bigg( \frac{4 \pi \rho^2 \tau^{-1}}{k^2 s} + \sigma_0 \tau^{-1} - i \sigma_Q \omega \bigg)\bigg(\tau^{-1} - i \omega \bigg) }{(\tau^{-1} + \gamma - i \omega)^2 + \omega_c^2} \nonumber \\
\sigma_{xy}(\omega) &=&  \frac{ \omega_c \bigg( \frac{4 \pi \rho^2 \tau^{-1}}{k^2 s} + (\sigma_0 + \sigma_Q) \tau^{-1} -2 i \sigma_Q \omega \bigg)}{(\tau^{-1} + \gamma - i \omega)^2 + \omega_c^2}
\label{sigma}
\end{eqnarray}
Similarly the thermoelectric conductivities are
\begin{eqnarray}
\alpha_{xx}(\omega) &=& \frac{\bigg( \frac{4 \pi \rho \tau^{-1}}{k^2} +  \frac{i \mu \sigma_Q \omega}{T} \bigg)\bigg(\tau^{-1} - i \omega \bigg) }{(\tau^{-1} + \gamma - i \omega)^2 + \omega_c^2} \nonumber \\
\alpha_{xy}(\omega) &=&  \frac{ \omega_c \bigg( \frac{4 \pi \rho \tau^{-1}}{k^2} + \frac{s  \sigma_Q \tau^{-1}}{\rho} + \frac{i \sigma_Q \omega}{\rho T}(\mu \rho- {sT}) \bigg)}{(\tau^{-1} + \gamma - i \omega)^2 + \omega_c^2}
\label{alpha}
\end{eqnarray}
and finally the heat conductivities are found to be
\begin{eqnarray}
\bar{\kappa}_{xx}(\omega) &=& \frac{\bigg( \frac{4 \pi s T \tau^{-1}}{k^2} -  \frac{i \mu^2 \sigma_Q \omega}{T} \bigg)\bigg(\tau^{-1} - i \omega \bigg) + \frac{\sigma_Q B^2}{T} }{(\tau^{-1} + \gamma - i \omega)^2 + \omega_c^2} \nonumber \\
\bar{\kappa}_{xy}(\omega) &=&  \frac{ \omega_c \bigg( \frac{4 \pi s T \tau^{-1}}{k^2} + \frac{2 i \omega \mu \sigma_Q s}{\rho} - \frac{\mu \sigma_Q \tau^{-1}}{\rho T}(2 sT + \mu \rho) \bigg)}{(\tau^{-1} + \gamma - i \omega)^2 + \omega_c^2}
\label{kappabar}
\end{eqnarray}
where all these expressions are accurate up to corrections of ${\cal O}(\varepsilon^2)$.
\para{}These formulae for the transport coefficients constitute one of the main novel results of this paper. As we emphasised earlier, beyond leading order in $\varepsilon$ they are different to those presented in \cite{nernst}. Nevertheless, we can perform perform various checks on their consistency. Firstly, in the absence of momentum relaxation (i.e $k=0$) they reduce to the usual results of relativistic hydrodynamics\footnote{These can be found in \cite{sduality,hallgreece} or by setting $\tau_{\mathrm{imp}}^{-1} =0$ in equation (3.37) of \cite{nernst}}. 
\paragraph{} Secondly, we can compare to the AC transport coefficients at $B=0$ that were previously derived for this model \cite{richblaise1, hydrous}. It is immediately clear that, in the limit $B \rightarrow 0$, we reproduce these existing results. However we can do much better. The four-dimensional action \eqn{action} exhibits an electromagnetic duality that fixes the transport coefficients of a dyonic black hole in terms of those of a purely electrically charged black hole (see Appendix~\ref{sec:appendix}). The fact that our expressions are consistent with this duality provides strong confirmation of our formulation of magnetohydrodynamics. 
\paragraph{}Finally, it is well known that within this model the DC transport coefficients can be expressed exactly in terms of horizon data. We have checked that, to the order we are working, the $\omega \rightarrow 0$ limit of our results agree with these formulae. Nevertheless, it is not immediately clear from the way we have written \eqn{sigma} \eqn{alpha} and \eqn{kappabar} in terms of $\omega_c, \gamma$ and $\tau^{-1}$ that this is the case. Whilst these are the natural variables that describe the time-evolution of the fluid flow, we will see in the next section that there is an alternative formulation of our hydrodynamics which is better suited to discussing the DC limit. 
\section{Hydrodynamics in the DC limit }
\label{sec:dclimit}
\paragraph{}In the last section we formulated our fluid-mechanics in the Landau-frame, such that the constitutive relation for the momentum density was proportional to the fluid velocity $T^{0 i} = (\epsilon + P) v_i$. This choice of frame was appropriate for studying the finite frequency conductivity because the Ward identity \eqn{wardidentity} includes time derivatives of $T^{0 i}$. 
\paragraph{}In the DC limit however, it is clear from \eqn{acward} that within our model it is the heat current, and not the momentum density, that plays the fundamental role\footnote{We do not have a deep understanding of why this is the case, but the same conclusion can be drawn for very general holographic models from the results of \cite{hydroaristos, hydroaristos2, hydrolucas}.}. As we remarked earlier, there is an ambiguity in the definition of the fluid velocity within relativistic hydrodynamics. We are therefore free to reformulate our constitutive relations in a new frame by introducing a new fluid velocity $\bar{v}_i$ such that the heat current can be written
\be
 Q_{i} = s T  \bar{v}_i
 \label{secondheatcurrentconstit}  
 \ee
 In practice this means that we are defining our new velocity through the relation 
 \be
 \bar{v}_{i} = v_{i} - \frac{\mu \sigma_Q}{s T} \bigg( -\partial_{i} \mu + {\cal F}_{i j} v_j + \mu \frac{\partial_{i} T}{T} \bigg) - \frac{\mu^2 k^2 \sigma_Q}{4 \pi s T^2} v_i + \dots
 \ee
Inserting this into the constitutive relations then allows us to write the electrical current in terms of ${\bar v}_i$ as 
\be
J_{i} = \rho {\bar v}_{i} + \bigg(\frac{\epsilon + P}{s T} \bigg) \sigma_Q \bigg (-\partial_{i} \mu + {\cal F}_{i j} {\bar v}_j + \mu \frac{\partial_{i} T}{T} \bigg) +  \bigg(\frac{\epsilon + P}{s T} \bigg)\frac{\mu \sigma_Q k^2}{4 \pi T} {\bar v}_i + \dots
\label{secondcurrentconstit}
\ee
\para{}Finally, it will prove convenient to make one further manipulation of this equation. To do this we recall that, at any given order in hydrodynamics, the various derivative terms that can appear in the constitutive relations are not all independent \cite{kovtun}. Rather, they are related by the hydrodynamic constraint equations, which arise from considering the equations of motion for the currents at lower orders in the expansion. In particular, we can use the ${\cal  O}({\varepsilon^2})$ expression for the Ward identity
\be
(\epsilon + P) \partial_{t} {\bar v}_i + \rho \partial_{i} \mu + s \partial_{i} T = - \frac{k^2 s}{4 \pi} \bar{v}_{i} + \rho {\cal F}_{i j} \bar{v}_{j} 
\ee
to eliminate the ${\sim k^2}$ term in \eqn{secondcurrentconstit}. This gives
\begin{eqnarray}
J_i = \rho \bar{v_i} + \sigma_0 \bigg(-\partial_{i} \mu + {\cal F}_{i j} {\bar v}_{j}\bigg) - \mu \sigma_{0} \partial_t \bar{v}_{i} + \dots
\label{thirdcurrentconstit}
\end{eqnarray}
where we note that in this new formulation it is now
\be
\sigma_{0} = \frac{1}{16 \pi G_N}
\ee
as opposed to $\sigma_Q$ that appears naturally in the constitutive relation for the current. 
\subsection*{DC transport coefficients}
\para{}We emphasise that the constitutive relations \eqn{secondheatcurrentconstit} and \eqn{thirdcurrentconstit} are just as valid as those we presented in Section~\ref{sec:linearhydro}. We are simply exploiting the ambiguities of hydrodynamics to rewrite the theory in a new frame. The motivation for doing this, as will now become clear, is that with these new constitutive relations the structure of the DC limit is self-evident. Indeed, taking the $\omega \rightarrow 0$ we simply have
\begin{eqnarray}
J_i &=& \rho \bar{v_i} + \sigma_0(-\partial_{i} \mu + {\cal F}_{ i j} {\bar v}_j ) \nonumber  \\
Q_i &=& s T \bar{v_i}
\label{dcconstit}
\end{eqnarray}
and can extract the transport coefficients by noticing that in the DC limit the Ward identity \eqn{acward} implies the constraint
\be
\rho \partial_{i} \mu + s \partial_{i} T = -\frac{k^2}{4 \pi T} Q_i + {\cal F}_{i j} J_j
\label{dcward}
\ee
which can be solved to determine the fluid velocity ${\bar  v_i}$. We find that the complexified velocity ${\bar v}_{+} = {\bar v}_{x} + i {\bar v}_{y}$ is given by
\be
\bar{v}_{+} = - \frac{\rho - i \sigma_0 B}{\frac{k^2 s}{4 \pi} + \sigma_0 B^2 + i \rho B} \partial_{+} \mu - \frac{s}{\frac{k^2 s}{4 \pi} + \sigma_0 B^2 + i \rho B} \partial_{+} T
\label{dcfluidvel}
\ee
Inserting this expression into the DC constitutive relations \eqn{dcconstit} then allows us to read off the thermoelectric response coefficients as
\begin{eqnarray}
\sigma_{xx} &=& \frac{\frac{k^2 s}{4 \pi} \bigg(\rho^2 + \sigma_0^2 B^2 + \frac{\sigma_0 k^2 s}{4 \pi} \bigg)}{ \bigg( \frac{k^2 s}{4 \pi} + \sigma_0 B^2 \bigg)^2 + \rho^2 B^2} \;\;\;\;\;\;\;\;\;\; \sigma_{xy} = \frac{ \rho B \bigg(\rho^2 + \sigma_0^2 B^2 + \frac{2 \sigma_0 k^2 s}{4 \pi}\bigg)}{ \bigg( \frac{k^2 s}{4 \pi} + \sigma_0 B^2 \bigg)^2 + \rho^2 B^2}  \nonumber \\
%
\alpha_{xx} &=& \frac{\rho s \frac{k^2 s}{4 \pi} }{ \bigg( \frac{k^2 s}{4 \pi} + \sigma_0 B^2 \bigg)^2 + \rho^2 B^2} \;\;\;\;\;\;\;\;\;\; \alpha_{xy} = \frac{ s B \bigg(\rho^2 + \sigma_0^2 B^2 + \frac{\sigma_0 k^2 s}{4 \pi}\bigg)}{ \bigg( \frac{k^2 s}{4 \pi} + \sigma_0 B^2 \bigg)^2 + \rho^2 B^2}  \nonumber \\
%
%
\bar{\kappa}_{xx} &=& \frac{s^2 T \bigg(\frac{k^2 s}{4 \pi} + \sigma_0 B^2 \bigg) }{ \bigg( \frac{k^2 s}{4 \pi} + \sigma_0 B^2 \bigg)^2 + \rho^2 B^2} \;\;\;\;\;\;\;\;\;\; \bar{\kappa}_{xy} = \frac{ \rho B s^2 T}{ \bigg( \frac{k^2 s}{4 \pi} + \sigma_0 B^2 \bigg)^2 + \rho^2 B^2}  
\label{exactDCformulae}
\end{eqnarray}
\para{}To ${\cal O}(\varepsilon^0)$, these expressions agree with the $\omega \rightarrow 0$ limit of the results in Section~\ref{sec:linearhydro}. However, it is worth emphasising that this agreement arises in a quite non-trivial manner. It is tempting, say, to think that the $\rho^2 B^2$ term in the denominators of \eqn{exactDCformulae} is the same as the $\omega_c^2$ factor in \eqn{sigma}. Likewise one might try to associate $(\epsilon + P) \tau^{-1}$ with the various factors of $\frac{k^2 s}{4 \pi}$.  However, we have already seen that beyond leading order things are not so simple  - there are subleading corrections in \eqn{correctedgamma} and \eqn{correctedcyclo} which are crucial in showing that the $\omega \rightarrow 0$ limit of our AC expressions agrees with these formulae. 
\subsection*{Exact DC hydrodynamics}
\para{}Strictly speaking, we can only trust these expressions \eqn{dcconstit} for the DC limit of the constitutive relations to ${\cal O}(\varepsilon^2)$.  Likewise we have only evaluated the Ward identity up to ${\cal O}(\varepsilon^4)$. Nevertheless, the results they imply for the DC conductivities \eqn{exactDCformulae} are nothing other than the exact formulae of \cite{magnetous}. In other words, they are known to hold regardless of the strength of the magnetic field $B$ or the scalar source $k$. It is therefore natural to suggest that the constitutive relations \eqn{dcconstit} and the constraint arising from the Ward identity \eqn{dcward} will continue to hold exactly, i.e. to all orders in our derivative expansion\footnote{However, we certainly should expect that at higher orders in the derivative expansion we will see additional finite $\omega$ corrections to \eqn{acward} and  \eqn{secondcurrentconstit}.}.  
\para{}These observations are deeply connected to a beautiful recent paper by Donos and Gauntlett \cite{hydroaristos} (similar ideas have been developed in \cite{hydroaristos2,hydrolucas,hydrolucas2}.). There it was shown that the DC conductivity of quite general holographic models can be understood from solving the forced Navier-Stokes equations for a fluid living on the black hole horizon. Remarkably this was an exact description, achieved without the need to take any sort of hydrodynamical limit.  
\para{}The DC constitutive relations that we have derived \eqn{dcconstit} for the boundary quantum field theory are, for our model, just the same as the exact constitutive relation of the horizon fluid in\footnote{For inhomogeneous models the relationship between the boundary theory and the horizon physics is expected to be more complicated.} \cite{hydroaristos,hydroaristos2}. Similarly, our expression for the fluid velocity, \eqn{dcfluidvel}, is identical to that obtained by solving the Navier-Stokes equation on the horizon. 
%
%
What we have demonstrated in this section is how these equations, and hence the formulae \eqn{exactDCformulae}, naturally arise in the boundary theory through reformulating our hydrodynamics in terms of a new fluid velocity ${\bar v_i}$.  

\acknowledgments
I am grateful to Aristomenis Donos and David Tong for many useful conversations and comments on a draft of this manuscript. I am funded through a Junior Research Fellowship at Churchill College, Cambridge. This work was supported in part by the European Research Council under the European UnionÕs Seventh Framework Programme (FP7/2007-2013), ERC Grant agreement STG 279943, Strongly Coupled Systems.

\appendix
\section{Electromagnetic duality} 
\label{sec:appendix}
\para{}In this appendix we set $16 \pi G_N =1$ for convenience. Then the four dimensional Einstein-Maxwell-Scalar action \eqn{action} exhibits an electromagnetic duality upon rotating the field strength $F_{M N}$ into its Hodge dual $\sqrt{-g} \epsilon_{M N P Q} F^{P Q}$. This symmetry allows us to map the electrically charged \RN\ black hole (\eqn{blackhole} with $u^{\mu} = (1,0,0)$) into a dyonic black hole.
\para{}Additionally, the duality acts on the perturbations of the black hole by rotating the current $J_+$ into the electric field $i E_{+}
$\cite{sduality,quantumcritical,dyonic}. As such it can be used to relate the transport coefficients of the dyonic black hole to those of a purely electrically charged background. In particular, letting $\sigma_+(\rho, B) = \sigma_{xx}(\rho,B) - i \sigma_{xy}(\rho,B)$ denote the conductivity of a dyonic black brane with charges $(\rho, B)$ we have the relation
\be
\sigma_{+}(\rho, B) = \frac{i \sigma_{+}(\sqrt{\rho^2 + B^2}, 0)\mathrm{cos}\theta - \mathrm{sin}\theta}{i \mathrm{cos}\theta - \sigma_{+}(\sqrt{q^2 + B^2}, 0) \mathrm{sin}\theta}
\ee
where $\mathrm{tan}\theta = B/\rho$. Similarly, there are corresponding expressions for the thermoelectric
\be
\alpha_{+}(\rho, B) =  \bigg( \mathrm{cos}\theta  - i \sigma_{+}(\rho, B) \mathrm{sin}\theta \bigg) \alpha_{+}(\sqrt{\rho^2 + B^2}, 0)
\ee
and heat conductivities
\be
\bar{\kappa}_{+}(\rho, B) = \bar{\kappa}(\sqrt{\rho^2 + B^2}, 0) - i T \alpha_+(\sqrt{\rho^2 + B^2}, 0) \alpha_+(\rho ,B) \mathrm{sin}\theta
\ee
Note that this rotation completely determines the low frequency transport coefficients (to ${\cal O}(\varepsilon^0)$) in terms of those previously calculated for the electrically charged black brane in \cite{richblaise1, hydrous}. We have checked that our expressions \eqn{sigma} \eqn{alpha} \eqn{kappabar} satisfy these relations.

\end{document}